\documentstyle[aps,prl,multicol,epsf,rotate]{revtex}

\begin{document}

\title{In-plane fluxon in layered superconductors with arbitrary number of layers}
\author{V.M.Krasnov}
\address{Department of Microelectronics and Nanoscience, Chalmers University of\\
Technology, S-41296 G\"oteborg, Sweden \\ and Institute of Solid
State Physics, 142432 Chernogolovka, Russia}
\date{\today }
\maketitle

\begin{abstract}
I derive an analytic solution for the in-plane vortex (fluxon) in
layered superconductors and stacked Josephson junctions with
arbitrary number of layers and interlayer coupling. For the stack
of $N$ junctions, the fluxon is characterized by $N$ spatial
components. The variation of the fluxon shape with $N$ is studied
analytically and numerically for $N=2,3,4,...600$. It is shown
that the fluxon shape strongly depends on $N$. With increasing
$N$, phase/current and magnetic field of the fluxon decouple from
each other. For comparison with real high-$T_c$ superconducting
samples, large scale numerical simulations with up to 600 SJJ's
and with in-plane length up to 4000 $\lambda_J$, are presented.
\end{abstract}

\begin{multicols}{2}

Strongly anisotropic layered superconductors can be considered as
stackes of Josephson junctions (SJJ's). Properties of SJJ's can be
qualitatively different both from that of bulk superconductors and
single Josephson junctions (JJ's). This is reflected in the
structure of in-plane vortex (fluxon). A knowledge of fluxon
structure is necessary for understanding of transport and magnetic
properties of layered superconductors. From the application point
of view, it is particularly interesting to consider structures
with a finite number of strongly coupled layers, which is the case
for high-$T_c$ (HTSC) mesas and artificial superconducting
multilayers, see e.g. Ref.\cite{Compar} and references therein.

So far there is no fluxon solution for SJJ's with arbitrary number
of junctions, $N$. Approximate solutions are available only in two
limiting cases: $N=\infty$ strongly coupled SJJ's\cite{Clem2}; and
$N=2$ for weakly coupled\cite{Malomed} or
arbitrary\cite{Modes,Fluxon1} SJJ's. The existence of isolated
fluxon\cite{Chush}, as well as the fluxon shape are still a matter
of controversy. For $N=\infty$, the fluxon can be characterized by
two length scales along the layers ($ab$-plane)\cite{Clem2}: the
Josephson penetration depth, $\lambda _J$, and the effective
London penetration depth, $\lambda _c$. Somewhat different
coefficients were derived in Ref.\cite{Koshel}, based on
simulations for $N=15$. In general, different conclusions about
the fluxon size could be made from available numerical simulations
for different $N$: short scale $\sim \lambda _J$ for $N$=7,
Ref.\cite{SBP}, and $N$=19, Ref.\cite {Klein1}; and long scale
$\sim \lambda _c$ for $N$=50, Ref.\cite{Shaf}. Long scale, $\sim
\lambda _c$, variation of $B$ for $N>1000$, is deduced from
experimental observations of the fluxon in
HTSC\cite{Kirtley,VDMarel}. For Bi and Tl-based HTSC, the ratio
$\lambda _c/\lambda _J$ can be well above 100 and it is not quite
clear which of those lengths, if any, represents the effective
Josephson penetration depth, $\lambda _{J(eff)} $. It is important
to know $\lambda _{J(eff)}$, since the behavior of SJJ's changes
drastically when the in-plane length, $L$, becomes longer than
$\lambda _{J(eff)}$. For example, in long strongly coupled SJJ's
multiple quasi-equilibrium fluxon modes\cite{Modes,Submodes}
exist, which result in anomalous perpendicular transport
properties, observed both for HTSC mesas and artificial Nb/Cu
multilayers\cite {Compar}.

In this paper I derive an approximate fluxon solution, which is
valid for any finite number of SJJ's and arbitrary electromagnetic
interlayer coupling. For the stack of $N$ junctions, the fluxon
consists of $N$ components with different characteristic lengths.
Variation of the fluxon shape is studied analytically and verified
numerically for $N=2,3,4,...600$ and for different interlayer
coupling. It is shown that fluxon structure strongly depends on
$N$. With increasing $N$, phase/current and magnetic field of the
fluxon decouple from each other. For quantitative comparison with
HTSC, large scale numerical simulations for $N$ up to $600$ and
$L$ up to 4000 $\lambda _J$ are presented.

First, I will briefly describe the formalism used in calculations.
Phase differences, $\varphi _i$, are described by the coupled
sine-Gordon equation (CSGE)\cite{SBP}:

\begin{equation}
{\bf \varphi }^{\prime \prime }={\bf A\cdot j}_c sin({\bf
\varphi}),  \label{Eq.1}
\end{equation}

where ${\bf A}$ is a tridiagonal matrix: $A_{i,i\mp 1}=-S_i/\Lambda _1$, $%
A_{i,i}=\Lambda _i/\Lambda _1$, $\Lambda _i$=$t_i+\lambda
_{Si}coth\left( \frac{d_i}{\lambda _{Si}}\right)+\lambda
_{Si+1}coth\left( \frac{d_{i+1}}{\lambda _{Si+1}}\right)$,
$S_i$=$\lambda _{Si}cosech\left( \frac{d_i}{\lambda _{Si}}\right)
$, $j_{ci}=J_{ci}/J_{c1}$ - the normalized critical current
density, $t_i$ - the thickness of the tunnel barrier, $d_i$ and
$\lambda _{Si}$ - the thickness and London penetration depth of
superconducting layers, $s_i=t_i+d_i$ - the interlayer spacing.
The subscript $i$ represents junction number. Space is normalized
to $\lambda _{J1}$ of a single JJ 1. For more details see
Refs.\cite{Modes,Fluxon2}. Eq.(1) should be supplemented by
boundary conditions,

\[
{\bf \varphi }^{\prime }\left( x=0,L\right) =\frac{2H\Lambda _i^{*}}{%
H_0\Lambda _1},
\]

where $\Lambda _i^{*}=\Lambda _i-S_i-S_{i+1}$ and $H_0=\frac{\Phi
_0}{\pi \lambda _{J1}\Lambda_1}$. Magnetic induction in the stack
is given by:

\begin{equation}
{\bf B}=\frac{H_0}2{\bf A}^{-1}{\bf \varphi }^{\prime }.
\label{Eq.2}
\end{equation}

For $d/\lambda _s\ll 1$, equations analogous to Eq.(1) are
obtained from Lawrence-Doniach (LD) model\cite{Clem2,Bul1}. The
LD-CSGE conversion is
shown in Table 1, along with estimations for Bi2212 HTSC: $d$=3 \AA , $t$%
=12 \AA , $J_c$=500-1000 A/cm$^2$, $\lambda _s$=750-1000 \AA .

For single JJ, the phase difference of the fluxon is:

\begin{equation}
F\left( \lambda _J\right) =4arctan\left( e^{x/\lambda _J}\right) .
\label{Eq.3}
\end{equation}

Both phase/current and magnetic induction are characterized by
$\lambda _J\equiv \lambda _J$(CSGE).

For SJJ's with arbitrary $N$ there is still no fluxon
solution. The problem with solving CSGE is the coupling of nonlinear $%
sin(\varphi _i)$ terms in the right-hand side of Eq.(1). To
decouple CSGE, I extend the approach of Ref.\cite{Fluxon1}. Let
$i_0$ be the number of JJ in which the fluxon is placed. By
diagonalizing matrix ${\bf A}$, Eq.(1) can be transformed to

\begin{equation}
\lambda _m^2F_m^{\prime \prime }=sin(F_m)+Er_m. \label{Eq.4}
\end{equation}

Here

\begin{equation}
F_m=\varphi _{i_0}+\sum \kappa _{m,i}\varphi _{i\neq i_0},
\label{Eq.5}
\end{equation}

$Er_m=sin(\varphi _{i_0})+\sum \kappa _isin(\varphi _{i\neq i_0})
- sin(\varphi _{i_0}+\sum \kappa _i\varphi _{i\neq i_0}) $, and
characteristic lengths, $\lambda _m^{-2}$ and coefficients $\kappa
_{m,i}$ are given by eigenvalues and eigenvectors of ${\bf A}$,
respectively. Let's observe that phase differences in JJ's not
containing the fluxon, $i \neq i_0$, are small, see Figs. 1,2.
Moreover $\varphi _{i\neq i_0}=0$ for $x=0,\pm \infty $.
Therefore, $Er_m$ have a form of small ripple around zero value
and vanish both inside and far from the fluxon center. In the
first approximation we may neglect $Er_m$ terms in the right hand
side of Eq.(4), which is essentially similar to partial
linearization of Eq.(1) with respect to $\varphi _{i\neq i_0}$.
Then we obtain a set of decoupled sine-Gordon equations for $F_m$.
I should note, that the diagonalization procedure used for
decoupling of CSGE minimizes the error functions, $Er_m$ far from
the center, when $\varphi_{i} \leftarrow 0$. Indeed, it can be
easily shown that in this case $Er_m \sim \varphi_i^3$, while for
any other linear combination of Eq.(1) it would be $\sim \varphi_i$.

Finally, inverting Eq.(5) we obtain the desired appoximate solution:

\begin{equation}
{\bf \varphi }={\bf K}^{-1}F_m(\lambda _m),  \label{Eq.6}
\end{equation}

where ${\bf K}$ is the $N\times N$ matrix with elements equal to
$\kappa _{m,i}$. From Eq.(6) it follows that the fluxon consists
of $N$ solitonic components, $F_m(\lambda _m)$, with
$N$ characteristic lengths, $\lambda _m$, (not just two length as in Ref.\cite{Clem2}).

Below I will consider identical SJJ's. In this case,
eigenvalues/vectors are given by\cite{SakPed},

\begin{eqnarray}
\lambda _m &=&\left[ 1+2Scos\left( \frac{\pi m}{N+1}\right)
\right] ^{-1/2},\ m=1,2, \ldots N,  \eqnum{7 a}  \label{Eq.7a} \\
\kappa_{m,i} &=&\left( -1\right) ^{i-i_0}\frac{sin\left[ \pi
mi/\left( N+1\right) \right] }{sin\left[ \pi mi_0/\left(
N+1\right) \right] }.  \eqnum{7 b} \label{Eq.7b}
\end{eqnarray}

Here $S=S_i/\Lambda _i$ is the coupling parameter, $S=0\div 0.5$.

\begin{figure}
\noindent
\begin{minipage}{0.48\textwidth}
\epsfxsize=0.9\hsize \centerline{ \epsfbox{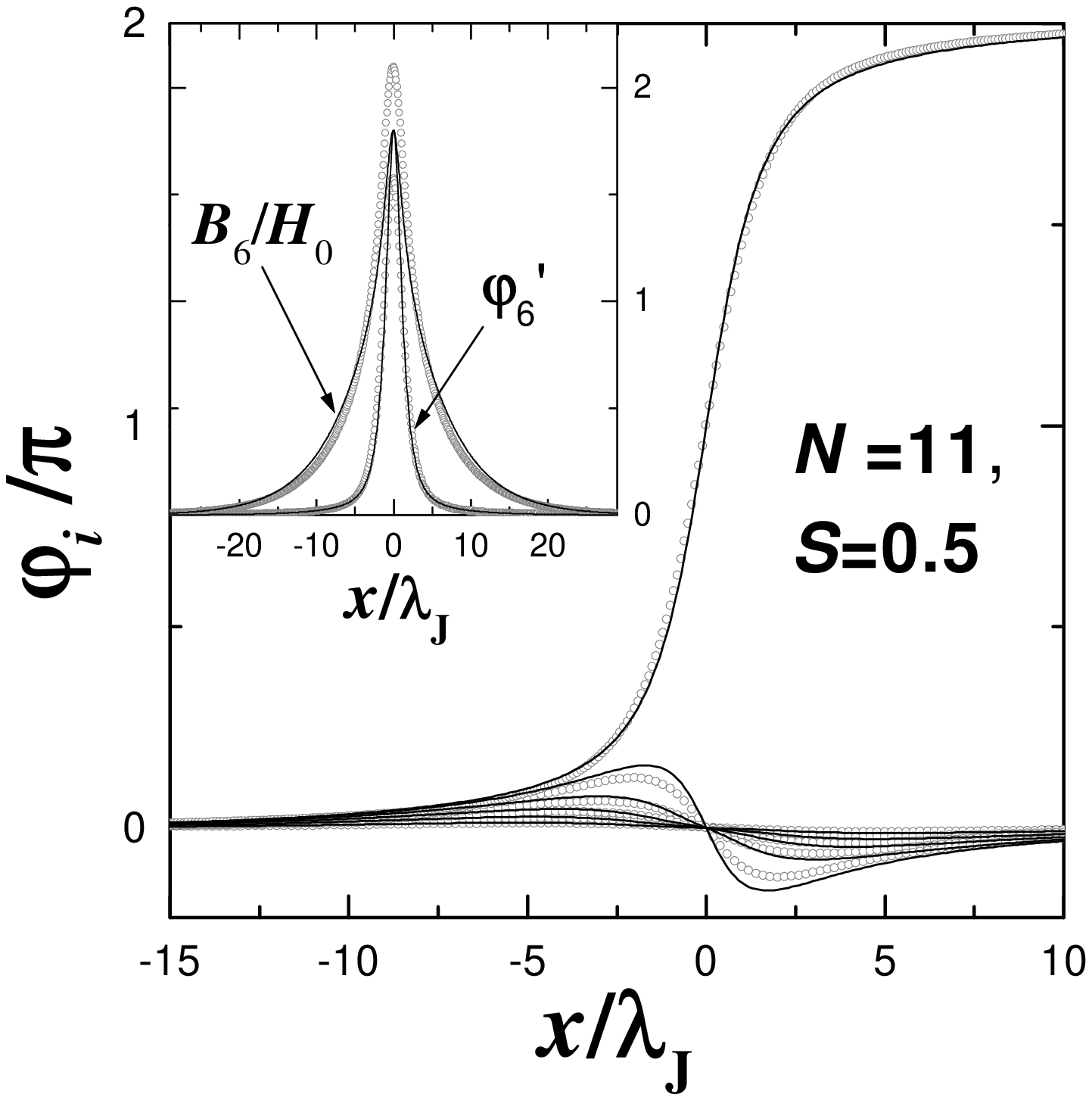} }
\vspace*{6pt} \caption{ Phase differences in the fluxon for $N$=11
SJJ's. Circles and lines represent numerical and analytic
solutions, Eq.(6), respectively. Inset shows variation of magnetic
induction and the derivative of phase in the central JJ. }
\label{fig1}
\end{minipage}
\end{figure}

For SJJ's with odd $N$ and fluxon in the middle junction,
$i_0=N/2$, the number of components, $F_m$, reduces to
$n=(N+1)/2$, due to a symmetry relation, $\varphi _{i_0-j}=\varphi
_{i_0+j}$. Then the solution becomes particularly simple:

\begin{eqnarray}
\varphi _{i_0} &=&\frac {2}{N+1}\sum F_m\text{ }\left( m=1,3,
\ldots N\right), \eqnum{8 a} \label{Eq.8a} \\ B_{i_0} &=
&\frac{H_0}{N+1}\sum \lambda _m^2F_m^{\prime }. \eqnum{8 b}
\label{Eq.8b}
\end{eqnarray}

Here are examples of fluxon solutions for small $N$:

%
%
\[
N=3:\left\{
\begin{array}{c}
\varphi _2=\frac{F_1+F_3}2,\lambda _1=\lambda _J\left( 1+\sqrt{2}S\right)
^{-1/2} \\
\varphi _{1,3}=\frac{-F_1+F_3}{2\sqrt{2}},\lambda _3=\lambda _J\left( 1-%
\sqrt{2}S\right) ^{-1/2}
\end{array}
\right.
\]

\[
N=5:\left\{
\begin{array}{c}
\varphi _3=\frac{F_1+F_3+F_5}3,\lambda _1=\lambda _J\left( 1+\sqrt{3}%
S\right) ^{-1/2} \\
\varphi _{2,4}=\frac{-F_1+F_5}{2\sqrt{3}},\lambda _3=\lambda _J \\
\varphi _{1,5}=\frac{F_1-2F_3+F_5}6,\lambda _5=\lambda _J\left( 1-\sqrt{3}%
S\right) ^{-1/2}
\end{array}
\right.
\]

Fig.1 shows distribution of phase differences for a single fluxon
in SJJ's with $N$=11, obtained by numerical simulation (circles)
and from analytic solution, Eq.(6), (solid lines). Parameters of
SJJ's are typical for Bi2212. Good agreement between analytic and
numerical solutions is seen. Inset shows spatial variation of $B$
and the derivative $\varphi ^{\prime }$ in the central JJ,
$i_0$=6. From inset it is seen that unlike single JJ, length
scales for variation of $\varphi $ and $B$ are different: $\varphi
$ varies at distances $\lambda(0) \simeq 1.3\lambda _J$, while the
characteristic length for variation of $B$ at large distances is
$\lambda(\infty) \simeq 5.4\lambda _J\simeq \lambda _{11}$, see
Eq.(7a). I should

\begin{figure}
\noindent
\begin{minipage}{0.48\textwidth}
\epsfxsize=0.9\hsize \centerline{ \epsfbox{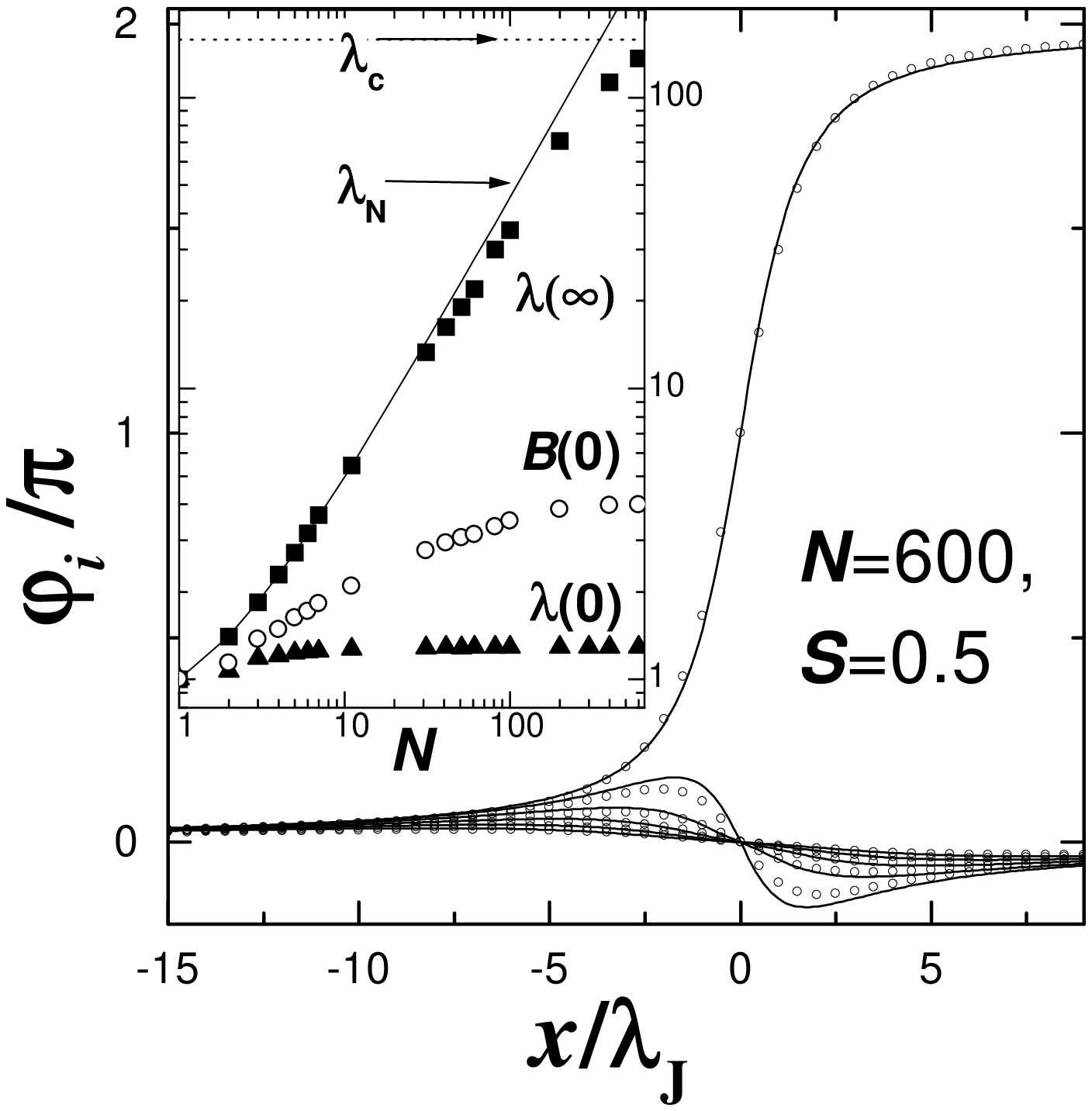} }
\vspace*{6pt} \caption{Phase differences in six central junctions
for $N$=600 SJJ's. Circles and lines represent numerical and
analytic solutions, Eq.(6), respectively. Inset shows variation of
characteristic parameters of the fluxon vs. the number of SJJ's. }
\label{fig2}
\end{minipage}
\end{figure}

\noindent emphasize, that the the overall fluxon shape is not
described by the two scales, $\lambda(0)$ and $\lambda(\infty)$,
but those two just represent asymptotic behavior at $x \rightarrow
0$ and $x \rightarrow \infty $, respectively. The appearance of
different length scales for $\varphi $ and $B$ can be understood
from Eq.(8). According to Eq.(8 a),

\begin{equation}
\lambda(0)=\frac 2{\varphi _{i_0}^{\prime }(0)}\simeq \frac n{\sum
\lambda _m^{-1}}, \eqnum{9} \label{Eq.9}
\end{equation}

and phase variation at $x \rightarrow 0$, is dominated by the
shortest length $\lambda _1 \sim \lambda _J$. On the other hand,
from Eq.(8 b), variation of $B$ is dominated by the longest
length, $\lambda _N$, which according to Eq.(7 a) increases
rapidly with $N$ for strongly coupled SJJ's, $S = 0.5$.

In order to study the fluxon shape in HTSC, it is necessary to
consider SJJ's with $N\gg \lambda _{ab}/s\simeq 100$ and $L\gg
\lambda _c\simeq 160\lambda _J$, see Table 1. Figs. 2,3 represent
results of large scale numerical simulations for $N$=600 SJJ's
with $L$=4000$\lambda _J$ and with parameters typical for Bi2212.
In Fig. 2, phase difference in six central JJ's, $i=300\div 305$,
are shown. Good agreement between numerical (circles) and analytic
(solid lines) solutions is seen.

Fig. 3 a) shows distribution of $B_i(x,z=const)$ along the $ab$-plane for
central junction, $i_0$=300, and for $i$=310. In inset, spatial
distributions of magnetic induction in the central junction,
$B_{300}$, obtained from numerical simulation (solid line),
analytic solution, Eq.(6) (dashed gray line), and from
Ref.\cite{Clem2} (dashed-dotted line), are compared. It is seen
that far from the core, numerical and analytic solutions perfectly
coincide. Such agreement was observed for all studied $N$ and $S$.
Dashed line in Fig. 3 a) represents the derivative
$d/dx(\varphi _{300})$ in the central junction. The dramatic
discrepancy in variation of $B_{i_0}$ and $\varphi _{i_0}^{\prime
}$ is seen.
Fig. 3 b) shows distribution of $B(z,x=const)$
along the $c$-axis for the central cross-section, $x$=0, and
for $x$=10$\lambda _J$. Inset represents a colour plot of $B(x,z)$
in the center of the fluxon.
It is seen that the most spectacular
feature of the fluxon is a sharp peak $B(0,0)$, representing the
''Josephson core'' of the fluxon $\sim \lambda _J\times s$ in
$ab$-plane and $c$-axis, respectively. For comparison, top axes in Figs. 3 a) and
b) are normalized to $\lambda _c$ and $\lambda _{ab}$,
respectively.

Inset in Fig. 2 summarizes the variation of fluxon shape for
different number of SJJ's. Open circles represent
maximum value of magnetic induction in the center of the fluxon,
$B(0)$. It increases with $N$ and saturates at $B(0)\simeq 4H_0$
for $N>\lambda _{ab}/s$. Solid triangles represent numerically
obtained $\lambda(0)$,
which determines variation of phase/current in the fluxon
core. In agreement with Eq.(9), $\lambda(0)$ increases only slightly with
$N$ and saturates at $\simeq 1.3\lambda _J$ for $N>10$. On the
other hand, far from the core the effective magnetic penetration
depth, $\lambda(\infty)$ (solid squares) along and across (not
shown) the layers increase dramatically with $N$. For $N<\lambda
_{ab}/s$, $\lambda(\infty)$ is determined by the largest
eigenvalue, $\lambda _N$, see Eq.(7 a), as shown by the solid line.
For $N>\lambda _{ab}/s$, $\lambda(\infty)$ becomes less
than $\lambda _N$ and shows a tendency for saturation at $\lambda
_c\simeq 160\lambda _J$, as shown by dotted line. In terms of the
analytic solution, Eqs.(6-8), this is due to contribution from the
rest of $N-1$ components.

In conclusion, approximate analytic solution for a single fluxon
in SJJ's with arbitrary number of junctions, $N$, and interlayer
coupling, $S$, was derived and verified by numerical simulations.
According to the solution, Eq.(6), the fluxon in the stack of $N$
junctions is described by $N$ characteristic lengths, Eq.(7 a).
For $N=2$, the validity of the solution was verified previously
both analytically \cite{Fluxon1} and numerically \cite{Fluxon2}.
Such multi-component fluxon could be qualitatively different from
that in a single JJ, Eq.(3). E.g. $B_2(0)$ can change sign in the
dynamic state\cite{Modes,Fluxon1}. Unusual fluxon shape certainly
has a strong impact on perpendicular transport properties of
SJJ's. For example, Lorentz contraction of the fluxon, and,
whence, the velocity matching behavior at the lowest Swihart
velocity should be substantially reduced\cite{Fluxon2} in SJJ's
with $N \gg1$ in comparison to single JJ's.

It was shown that the fluxon shape depends strongly on the number
of junctions in the stack. This resolves seeming contradictions
which appear when comparing simulations made for different $N$,
\cite{Koshel,SBP,Klein1,Shaf} with each other and with the
expression for $N=\infty$ \cite{Clem2}. For large $N$, the
obtained analytic solution asymptotically approaches the solution
for $N=\infty$ \cite{Clem2}, however, two characteristic length
scales, $\lambda _c$ and $\lambda _J$ appear due to nontrivial
averaging of all $N$ components. I should note, that even for
$N=600$ the agreement at large distances is not complete, see
inset in Fig. 3 a), since $\lambda(\infty)$ is still less than
$\lambda_c$, see inset in Fig. 2. On the other hand, the analytic
solution, Eq.(6), is in a perfect agreement with numerical
\end{multicols}

\begin{figure}
\noindent
\begin{minipage}{0.99 \textwidth}
\epsfxsize=0.95\hsize \centerline{ \epsfbox{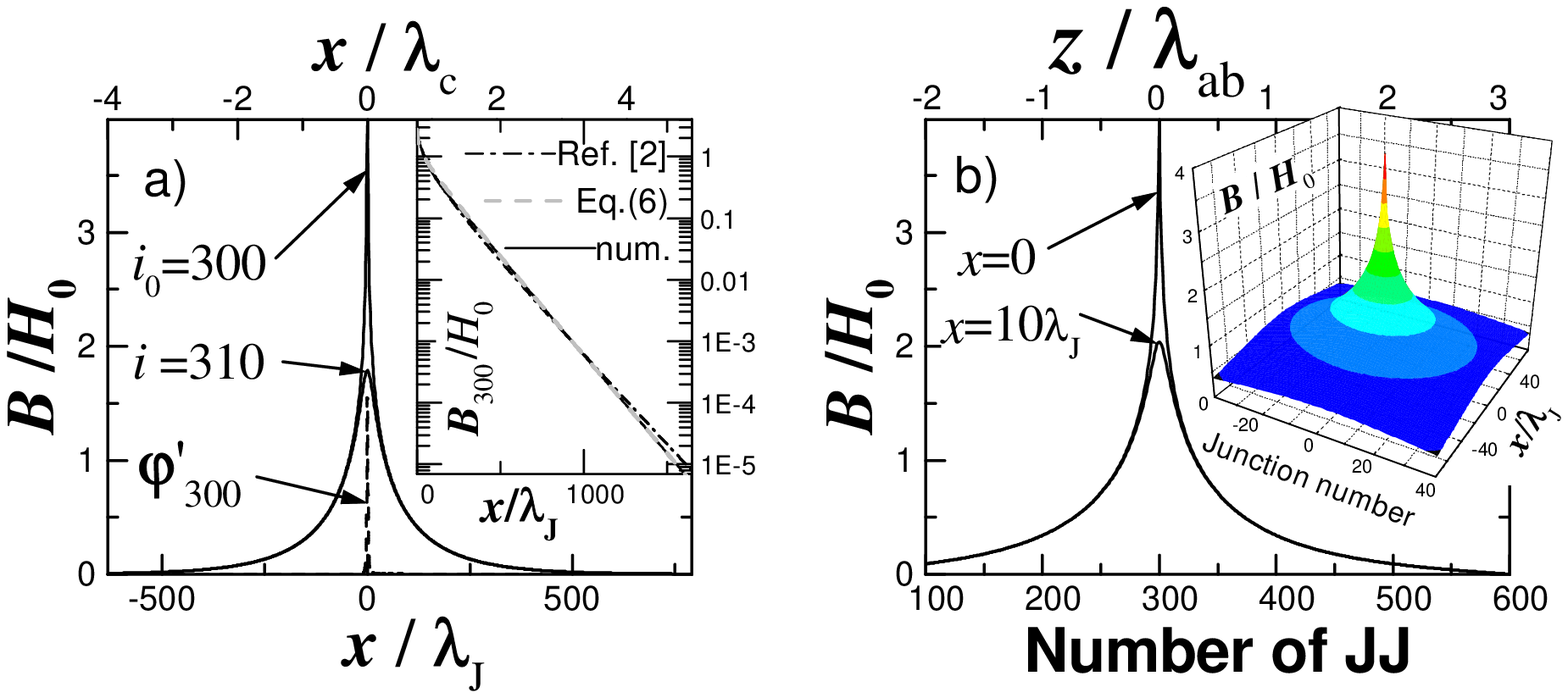} }
\caption{Numerically obtained magnetic field distribution of the
fluxon for $N$=600 SJJ's: a) $B(z=const)$ along the $ab$-plane.
Dashed line shows the derivative of phase in the central junction.
Inset represents comparison between numerical, analytic solution
and asymptotic expression for $N=\infty$ from Ref.[2] for
distribution of $B$ in the central junction. b) $B(x=const)$ along
the $c$-axis. Inset represents a colour plot of $B(x,z)$ in the
center of the fluxon.} \label{fig3}
\end{minipage}
\end{figure}

\begin{multicols}{2}

\noindent simulations, as shown by inset in Fig. 3 a).

An interesting property of the fluxon in strongly coupled SJJ's,
$S\simeq0.5$, is that with increasing $N$ the distribution of
phase/current decouple from magnetic field. Variation of phase is
almost independent of $N$ and occurs at distances $\simeq 1.3
\lambda _J$ along the planes, which represents the effective
Josephson penetration depth. On the other hand, magnetic induction
continues to vary at much larger distances, $\lambda(\infty) \sim
\lambda_N$. The decoupling is due to ineffective screening of
magnetic field by thin layers, $d/\lambda _s\ll 1$. From Eq.(7 a)
it is seen that in the case of weak coupling, $S \rightarrow 0$
(i.e. for thick layers), the splitting between the components
vanishes, $\lambda_m \rightarrow \lambda_J$, and the solution
returns to the form for single JJ, Eq.(3), without decoupling
between phase/current and magnetic field.

Finally, for comparison with HTSC, large scale numerical
simulations with $N=600$ and $L $=4000$\lambda _J$ were performed.
It was shown that the sharp peak $%
B(0,0)$ within the Josephson core is the most striking feature of the
fluxon. Direct experimental observations of the fluxon in HTSC\cite
{Kirtley,VDMarel} show no signature of such peak, probably due to
insufficient spatial resolution. The question, if it is possible to observe
the fluxon core experimentally is quite important since it may give a clue
to the nature of interlayer coupling in HTSC. From Fig. 3 it follows, that $%
B(0,0)\simeq 4H_0$, is small $\sim $ few Gauss. However, the
gradient of magnetic field could be as large as $10^7$ G/cm along
the $c$-axis. Therefore, it might be possible to observe the
fluxon by magnetic decoration method\cite{Dolan}, or other techniques
sensitive to the gradient of $B$.

The work was supported by INTAS 96-0452 and Swedish
Superconductivity Consortium.

\begin{table}[b]
\noindent
\begin{minipage}{0.47\textwidth}
\caption{Parameters of Lawrence-Doniach model in terms of CSGE and
estimations for Bi2212}
\begin{tabular}{|cccc|}
\multicolumn{1}{|c|}{LD} & \multicolumn{1}{c|}{$\lambda _{ab}$} &
\multicolumn{1}{c|}{$\lambda _c=\gamma \lambda _{ab}$} &
\multicolumn{1}{c|}{$\lambda _J=\gamma s $} \\ \hline
\multicolumn{1}{|c|}{CSGE} & \multicolumn{1}{c|}{$\lambda
_s\sqrt{\frac sd}$} & \multicolumn{1}{c|}{$\sqrt{\frac{\Phi
_0c}{8\pi ^2J_cs}}$} & \multicolumn{1}{c|}{$\sqrt{\frac{\Phi
_0cd}{16\pi ^2J_c\lambda _s^2}}=\frac{\lambda
_J(\text{LD})}{\sqrt{2}}$} \\ \hline \multicolumn{1}{|c|}{Bi2212}
& \multicolumn{1}{c|}{0.15-0.2 $\mu
m$} & \multicolumn{1}{c|}{130-184 $\mu m$} & \multicolumn{1}{c|}{$\lambda _J($%
CSGE$)=$0.6-1.1 $\mu m$} \\
\end{tabular}
\end{minipage}
\end{table}

\end{multicols}

\end{document}